\documentclass[reprint,amsmath,amssymb,aps,pre,showkeys]{revtex4-2}

\usepackage{graphicx}
\usepackage{dcolumn} 
\usepackage{upgreek}
\usepackage{bm}

\usepackage{mathtools}
\usepackage{cancel}
\usepackage{soul}
\usepackage{xcolor}
\usepackage{booktabs} 
\newcommand{\vb}[1]{{\mathbf #1} }

\begin{document}


\title{Competing chemical gradients change chemotactic dynamics and cell distribution} 

\author{Emiliano Perez Ipi\~na}
\affiliation{Department of Physics \& Astronomy, Johns Hopkins University, Baltimore, Maryland 21218, USA.}
\author{Brian A. Camley}%
\affiliation{Department of Physics \& Astronomy and Biophysics, Johns Hopkins University, Baltimore, Maryland 21218, USA.}


\date{\today}

\begin{abstract}
Cells are constantly exposed to diverse stimuli--chemical, mechanical, or electrical--that guide their movement. In physiological conditions, these signals often overlap, as seen during infections, where neutrophils and dendritic cells navigate through multiple chemotactic fields. How cells integrate and prioritize competing signals remains unclear. For instance, in the presence of opposing chemoattractant gradients, how do cells decide which direction to go? When should local signals dominate distant ones? A key factor in these processes is the precision with which cells sense each gradient, which depends non-monotonically on concentrations. Here, we study how gradient sensing accuracy shapes cell navigation in the presence of two distinct chemoattractant sources. We model cells as active random walkers that sense local gradients and combine these estimates to reorient their movement. Our results show that cells sensing multiple gradients can display a range of chemotactic behaviors, including anisotropic spatial patterns and varying degrees of confinement, depending on gradient shape and source location. The model also predicts cases where cells exhibit multistep navigation across sources or a hierarchical response toward one source, driven by disparities in their sensitivity to each chemoattractant. These findings highlight the role of gradient sensing in shaping spatial organization and navigation strategies in multi-field chemotaxis.
\end{abstract}

\maketitle

\section*{Introduction}

Directed cell migration plays a fundamental role in immune responses, tissue repair, development, and cancer progression. 
During these processes, cells navigate their local environments by responding to diverse cues, including chemical, mechanical, and electrical signals~\cite{shellard2020all,SenGupta2021-rk,lara2013directed}. 
These signals convey directional information through concentration gradients, extra-cellular fiber alignment, or electric field direction. 
Physiological environments frequently expose cells to multiple overlapping guidance cues \cite{zhao2009electrical,kolaczkowska2013neutrophil,petri2018neutrophil,moon2023cells}. 
For instance, electric fields and chemotactic gradients of chemokines and growth factors both guide fibroblast cells during wound healing~\cite{zhao2009electrical}. 
Similarly, during infections, neutrophils transmigrate across the endothelium and migrate toward infection sites by responding to multiple overlapping chemoattractant gradients, including \textit{N}-formyl-methionyl-leucyl-phenylalanine (fMLP), complement anaphylatoxin C5, leukotriene B$_4$ (LTB$_4$), and chemokines like interleukin-8 (IL-8) among others~\cite{kolaczkowska2013neutrophil,petri2018neutrophil}.
Likewise, dendritic cell migration within lymph nodes is guided by chemokine gradients of CCL19 and CCL21, with CCL19 being more potent at recruiting cells~\cite{liu2021dendritic}.  
These signals can act cooperatively, reinforcing migration in the same direction, or as competing cues, pulling cells toward different targets. How do cells integrate multiple guiding signals? How do they prioritize competing cues? Neutrophils placed between two opposing chemoattractant sources, IL-8 and $\mathrm{LTB}_4$, exhibit preferential migration toward the distant source~\cite{foxman1997multistep,foxman1999integrating,lin2005neutrophil}. Conversely, when cells were exposed to an ``end-target'' chemoattractant like fMLP alongside ``intermediate'' chemoattractants such as IL-8 or $\mathrm{LTB}_4$, they exhibited a hierarchical response to fMLP. Specifically, when closer to the fMLP source, cells migrated directly toward it. However, when initially closer to the ``intermediate'' chemoattractant source, they first moved toward the intermediate source before redirecting to fMLP, resulting in a stepwise migration across gradients~\cite{foxman1997multistep}. Experimental tracking of individual cells showed neutrophils exhibit oscillatory motion when exposed to opposing gradients of IL-8 and $\mathrm{LTB}_4$, but directed migration towards fMLP sources when this is combined with IL-8 and $\mathrm{LTB}_4$~\cite{byrne2014oscillatory}.
Similarly, mature dendritic cells demonstrate confinement when placed between two opposing chemoattractant sources~\cite{ricart2011dendritic}. When exposed to combinations of opposing CXCL12, CCL19, and CCL21 chemokine gradients, they migrate toward a central region and remain near a ``line of equistimulation'', where the chemotactic responses are balanced.

Several mechanisms have been proposed to explain how cells prioritize distant chemoattractant sources over closer ones. These include gradient sensing memory~\cite{foxman1999integrating,oelz2005multistep}, adaptation to chemoattractants via receptor desensitization~\cite{lin2008modeling,wu2011modeling}, and the inhibition of one chemoattractant by another~\cite{byrne2014oscillatory,heit2002intracellular}.
Here, we start with a different focus. Chemotaxis is a noisy process. Experiments and theory have established that instead of following a single gradient direction perfectly, cells have a distribution of orientations around the gradient direction, with the size of this distribution controlled by the signal-to-noise ratio (SNR) of each gradient~\cite{hu2010physical,van2007biased,fuller2010external,lim2018chemotaxis,segota2013high,amselem2012control}.
Sensing accuracy varies non-monotonically with chemoattractant concentration \cite{fuller2010external,lim2018chemotaxis,hu2010physical}: cells detect gradients most precisely at intermediate concentrations and less effectively at very low or very high concentrations. We explore if this effect explains why cells migrate toward distant sources.
We model cells sensing two types of chemoattractant as active random walkers that locally sense gradients and reorient their movement based on these estimates. We find a spectrum of chemotactic behaviors, including anisotropic spatial distributions, multistep navigation across sources, and hierarchical responses favoring one chemoattractant over the other. 
These behaviors emerge from the relative SNRs of the competing chemotactic gradients. {Our model explains many previously observed experimental phenomena solely by accounting for the accuracy of chemical gradient sensing, without requiring additional mechanisms such as those mentioned before.}

\begin{figure}[!htb]
    \centering
    \includegraphics[width=0.95\linewidth]{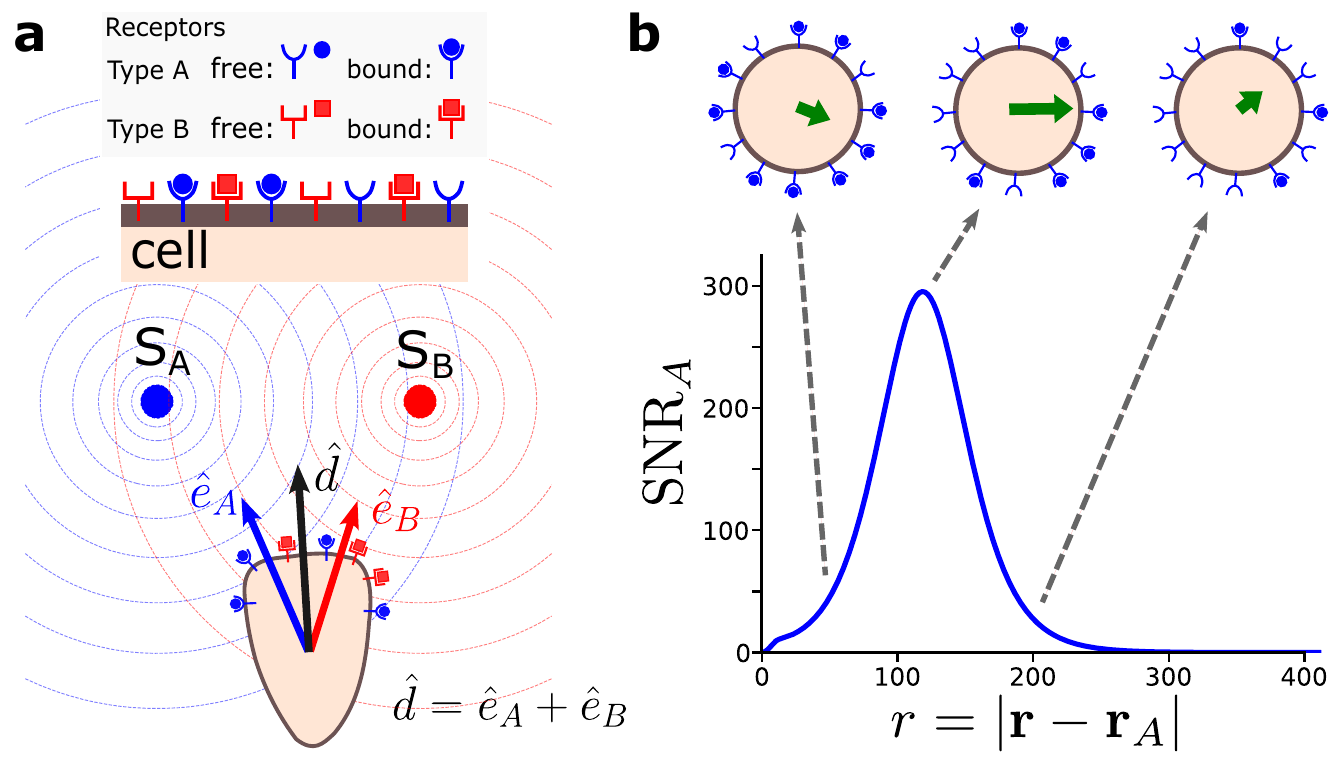}
    \caption{\textbf{a} Two sources, $S_A$ and $S_B$, secrete different chemoattractants, $A$ and $B$, respectively. Cells possess independent receptors of different types, each specifically binding to one of the chemoattractants.  \textbf{b} Signal-to-noise ratio (SNR) curve for chemoattractant $A$ as a function of the distance from the source. The SNR exhibits a non-monotonic shape, reaching its maximum sensitivity at $c_A(r^*) = K_D$. On top, a schematic representation of cells at different distances from the source. Both near and far from the source, receptor occupancy is either too high or too low, limiting the cell's ability to accurately measure the gradient.}
    \label{fig:figure1}
\end{figure}

\section*{Results}
\subsection*{Estimating cell migration direction in the presence of two chemoattractant sources.}
We consider two sources secreting different chemoattractants, labeled $A$ and $B$, Fig.~\ref{fig:figure1}. The chemoattractant concentration profile is described by the steady-state solution of a synthesis-diffusion-degradation (SDD) model, where molecules are released at the sources, diffuse with a diffusion coefficient $D_s$, and degrade at a rate $\gamma_s$. The concentration is given by:
\begin{equation}\label{eq:c_vs_r}
    c_s(\vb{r}) = S_{0_s} \frac{e^{-\vert \vb{r}-\vb{r}_s\vert/\lambda_s}}{\vert \vb{r}-\vb{r}_s\vert} \, f_s(\vert \vb{r}-\vb{r}_s\vert/\lambda_s,\epsilon),
\end{equation}
where $S_{0_s}$ is the source strength parameter that sets the concentration scale at the source location $\mathbf{r}_s$, $\lambda_s = \sqrt{D_s/\gamma_s}$ is the decay length of the gradient, and $s$ denotes the chemoattractant species $A$ or $B$. The function $f_s$ is a regularization factor introduced to prevent a divergence at $\vb{r} = \vb{r}_s$; $\epsilon=16\upmu\mathrm{m}^2$ sets the effective spatial extent of the source.
Details on the derivation of Eq.~(\ref{eq:c_vs_r}) can be found in Appendix~\ref{app:sources}.\\
We model cells as persistent random walkers with an additional alignment to the chemoattractant direction:
\begin{align}
    \dot{\vb{r}} &= v_0 \hat{\vb{e}}(\phi), \label{eq:prw_rdot} \\
    \dot{\phi} &= -\frac{1}{\tau} \sin(\phi - \hat{\phi}) + \sqrt{2D_\phi} \, \xi(t), \label{eq:prw_thetadot}
\end{align}
where $\phi$ is the direction of cell movement, $\hat{\vb{e}} = (\cos\phi, \sin\phi)^\mathrm{T}$ is the unit vector indicating the direction of motion, $\hat{\phi}$ is the estimated chemotactic direction (which we discuss later), $v_0$ is the cell speed, and $\tau$ is the characteristic time for alignment toward a given direction.
The second term in Eq.~(\ref{eq:prw_thetadot}) represents noise arising from fluctuations in cell polarization, accounting for the stochasticity of biochemical reactions regulating polarization, the finite number of molecules involved, and the dynamic remodeling of the cytoskeleton. The noise satisfies $\langle \xi(t) \rangle = 0$ and $\langle \xi(t)\xi(t') \rangle = \delta(t-t')$, with $D_\phi$ the angular diffusion coefficient that the cell's orientation would have in the absence of chemotactic alignment.\\

In this context of multiple sources, we assume that cells possess two receptor types, $A$ and $B$, each independently binding to their respective chemoattractant molecules. Consequently, cells independently estimate the direction of each local gradient $\phi_s$, and then integrate these estimates to determine their movement direction. We model this integration as the vector sum of the individually sensed directions:
\begin{equation}\label{eq:estimated_direction}
    \hat{\vb{d}} = \hat{\vb{e}}_A + \hat{\vb{e}}_B, 
\end{equation}
where $\hat{\vb{e}}_s = (\cos\hat{\phi}_s, \sin\hat{\phi}_s)^\mathrm{T}$ represents a unit vector in the estimated direction of each local gradient. This vector summation is supported by experimental observations and has been proposed in multiple studies including eukaryotic cells and bacteria~\cite{foxman1997multistep,foxman1999integrating,lin2005neutrophil,wu2011modeling,strauss1995analysis, raikwar2025phototactic}.

Cells sense and respond to chemotactic fields by estimating the local gradient through receptors distributed on their surface, which bind chemoattractant molecules. However, this estimation is fundamentally limited by the stochastic arrival and binding of chemoattractant molecules. Consequently, cells can only estimate the local orientation of a chemical gradient, $\phi_s$, with finite accuracy~\cite{berg1977physics,hu2010physical}, denoted by $\sigma_{\phi_s}$. 
We assume cells draw the estimated direction of each local gradient, $\hat{\phi}_s$, from a von Mises distribution, a generalization of the normal distribution for periodic variables, $\hat{\phi}_s(\vb{r}) \sim \mathcal{VM}(\phi_s(\vb{r}),\kappa_s(\vb{r}))$, with probability density function given by:
\begin{equation}\label{eq:phis_VMdist}
    p\left(\hat{\phi}_s|\phi_s,\kappa_s\right) = \frac{1}{2\pi I_0(\kappa_s)} \exp \left(\kappa_s\cos(\hat{\phi}_s - \phi_s)\right),
\end{equation}
where $I_0$ is the modified Bessel function of the first kind of order zero, $\phi_s(\vb{r})$ is the true gradient direction towards source $s$ from position $\vb{r}$. $\kappa_s$ determines the precision of gradient sensing -- a higher $\kappa_s$ corresponds to a more precise gradient orientation estimate. At large $\kappa_s$, the von Mises distribution limits to a normal distribution with variance $\sigma_{\phi_s}^{-2}=\kappa_s$.
To determine the gradient sensing error, $\sigma_{\phi_s}$, we follow the approach in~\cite{hu2010physical}. For a circular cell of radius $R_\mathrm{cell}$ with $n_s$ receptors uniformly distributed along its surface, each with a dissociation constant $K_{D_s}$, the error in estimating the gradient direction $\phi_s$ at a position where the concentration at the cell center is $c_s$ is given by $\sigma_{\phi_s}^2 = \frac{\sigma_{p_s}^2}{p_s^2}$, where the gradient steepness is defined as, $p_s = \frac{2 R_\mathrm{cell}}{c_s} \vert \nabla c_s \vert$, and its estimated variance is, $\sigma_{p_s}^2 = \frac{8(c_s + K_{D_s})^2}{n_s c_s K_{D_s}}$. Combining these expressions yields,
\begin{equation}\label{eq:sensing_error}
    \sigma_{\phi_s}^2 = \frac{2}{n_s\,R^2_\mathrm{cell}} \, \frac{\left( c_s + K_{D_s}\right)^2}{c_s\,K_{D_s}} \,\left(\frac{\vert \nabla c_s\vert}{c_s}\right)^{-2}.
\end{equation}
While \cite{hu2010physical} derived this form only for shallow gradients, we have previously found it to be a good approximation unless gradients are extraordinarily large, and expect it to capture the correct qualitative behavior \cite{hopkins2019leader}.We define the signal-to-noise ratio (SNR) as the inverse of $\sigma_{\phi_s}^2$,  $\mathrm{SNR}_s = \frac{p_s^2}{\sigma_{p_s}^2} = \sigma_{\phi_s}^{-2}$, and set $\kappa_s = \mathrm{SNR}_s$.
Finally, from Eq.~(\ref{eq:estimated_direction}), the overall estimated chemotactic direction is obtained as,
\begin{equation}\label{eq:estimated_direction_2}
    \hat{\phi} = \mathrm{atan2}\left(\sin(\hat{\phi}_{A}) + \sin(\hat{\phi}_B),\cos(\hat{\phi}_{A}) + \cos(\hat{\phi}_B) \right).
\end{equation}

It is important to note that $\hat{\phi}_s$ drawn from Eq.~(\ref{eq:phis_VMdist}) represents a snapshot measurement taken at a particular state of receptor binding occupancy, with a corresponding uncertainty error given by Eq.~(\ref{eq:sensing_error}). 
However, over time, cells perform new measurements and update their estimates of the gradient directions~\cite{berg1977physics,kaizu2014berg,perez2018fluctuations}.
To account for this,  we assume that cells perform discrete and synchronous measurements-- measure both gradients at the same time-- of $\phi_s$ every $\tau_\mathrm{sensing} = 1\,\mathrm{min}$. This is a reasonable approximation, as it aligns with the typical correlation times of the chemoattractant-receptor binding dynamics and is shorter than the characteristic timescale for cell displacement and reorientation, $\tau_\mathrm{sensing} < \tau$~\cite{ipina2022collective,camley2017cell}.

\subsection*{Combining two chemoattractant sources creates diverse chemotactic responses}
Here, we consider two sources positioned at $\vb{r}_{A} = -200\,\upmu\mathrm{m}\,\hat{\vb{x}}$ and $\vb{r}_{B} = 200\,\upmu\mathrm{m}\,\hat{\vb{x}}$. We initially assume both sources and chemoattractants have identical properties, so cells exhibit equal sensitivity to both chemoattractants.  
For simplicity, we omit the index $s$ from the gradient parameters. Unless explicitly stated otherwise, we use the notations $S_{0_s} = S_0$, $\lambda_s = \lambda$, $D_s = D$, $n_s = n$, $K_{D_s} = K_D$, and so forth.  

\begin{figure}[h]
    \centering
    \includegraphics[width=\linewidth]{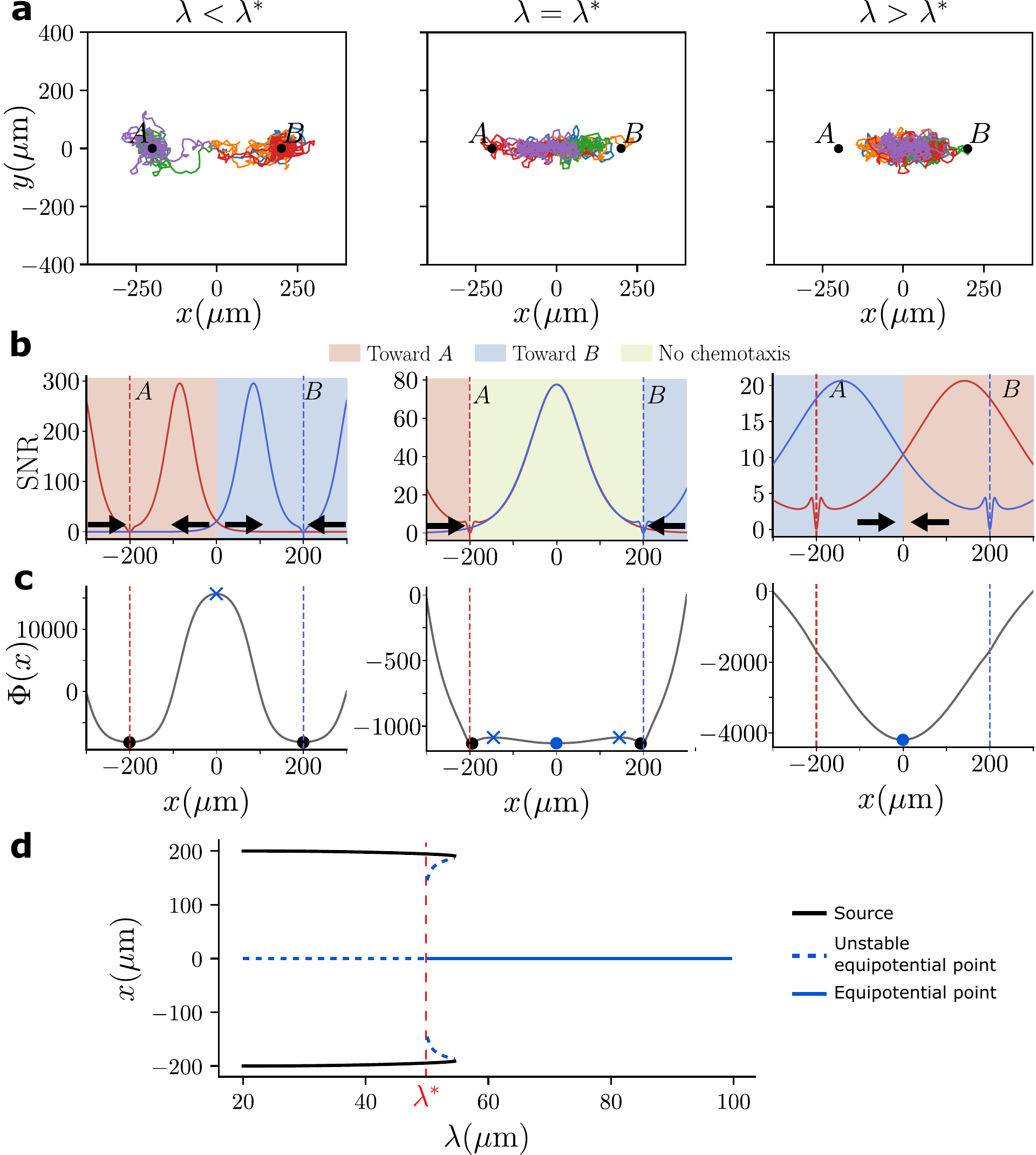}
    \caption{\textbf{(a)} Representative cell trajectories ($n=5$) for different values of $\lambda$ (left panel: $\lambda=50\upmu\mathrm{m}$, center panel: $\lambda^*=50\upmu\mathrm{m}$, right panel: $\lambda=100\upmu\mathrm{m}$), with initial positions centered between the two sources. \textbf{(b)} Signal-to-noise ratio (SNR) along the axis connecting the sources, denoted as $r$. The positions of sources A and B are marked by blue and red dashed lines, respectively. The blue and red solid lines represent $\mathrm{SNR}_A$ and $\mathrm{SNR}_B$. The red-shaded region indicates where cells are directed toward source A, the blue-shaded region corresponds to movement toward source B, and the green-shaded region marks areas where cells exhibit no chemotactic bias. \textbf{(c)} Chemotactic potential $\Phi(x)$ corresponding to the plots above in (\textbf{b}). Dots and crosses denote stable and unstable fixed points, respectively. Blue markers indicate equipotential fixed points where $\mathrm{SNR}_A = \mathrm{SNR}_B$. \textbf{(d)} Bifurcation diagram illustrating the transitions in chemotactic response as $\lambda$ varies. A pitchfork bifurcation occurs at $\lambda = \lambda^*$, where the equipotential fixed points transition from unstable to stable -- see Appendix~\ref{app:bifurcation} for details. Model parameters are shown in Table~\ref{tab:parameters}.}
    \label{fig:figure2}
\end{figure}

To investigate the chemotactic response of cells in overlapping gradients, we initialize them at $\vb{r}(t=0) = \vb{0}$, centrally positioned between the two sources. Their movement is then determined by numerically integrating Eqs.~(\ref{eq:prw_rdot})-(\ref{eq:prw_thetadot}), as detailed in Appendix~\ref{app:num_sim}.
We find that at small values of the decay length $\lambda$, cells will spontaneously break symmetry and migrate toward one of the two sources, then most likely remain near that source, (Fig.~\ref{fig:figure2}\textbf{a}). However, at larger $\lambda$, cells are confined to a region between the two sources.
These different outcomes can be attributed to the spatial distribution of the cell's gradient sensing signal-to-noise ratio ($\mathrm{SNR}_s$). 
At any given point in space, cells tend to align on average toward the source with the highest $\mathrm{SNR}_s$. 
Specifically, if $\mathrm{SNR}_A(\mathbf{r}) > \mathrm{SNR}_B(\mathbf{r})$, cells estimate the direction of field A with greater accuracy than field B, and vice versa. 
Receptor saturation reduces the cells' sensitivity near the sources when the chemoattractant concentration at the source exceeds the dissociation constant, $K_D$, i.e., $ c_s(\mathbf{r}_s) > K_D$, (Fig.~\ref{fig:figure1}\textbf{b}). 
Consequently, the $\mathrm{SNR}_s$ profile is non-monotonic: it increases with distance from the source up to a peak, after which it decreases. 
The relative positions of the $\mathrm{SNR}_s$ peaks along the axis connecting both sources, controlled by $\lambda$, determine the observed outcomes (Fig.~\ref{fig:figure2}\textbf{b}). 
For small $\lambda$, $\mathrm{SNR}_s$ peaks are proximal to their respective sources, leading cells to migrate toward the nearest source (left panel, Fig.\ref{fig:figure2}\textbf{b}). 
As $\lambda$ increases, the peaks shift away from the sources. For sufficiently large $\lambda$, the peaks cross and move closer to the opposite source. 
Under these conditions, cells move away from the closer source and toward the farther source (right panel, Fig.~\ref{fig:figure2}\textbf{b}). However, as the cells approach the initially distant source, the roles reverse: the previously distant source becomes the near source, and the near source becomes distant, prompting the cells to turn around. This dynamic causes the cells to oscillate back and forth between the two sources.
For an intermediate value of $\lambda$, denoted as $\lambda^*$, the peaks overlap precisely in between the sources -- see Appendix~\ref{app:bifurcation} for how $\lambda^*$ was obtained. 
In this scenario, between the two sources, $\mathrm{SNR}_A = \mathrm{SNR}_B$, indicating that cells sense both fields equally well. 
However, since the two fields oppose each other in direction, their effects cancel out, and cells exhibit no directional bias, resulting in the absence of chemotaxis (center panel, Fig.~\ref{fig:figure2}\textbf{b}).
Thus, cells spread between the sources, exhibiting a broader dispersion compared to the $\lambda>\lambda^*$ case.
\\

The regions where the responses to the chemoattractants balance and cancel out ($\mathrm{SNR}_A = \mathrm{SNR}_B$), and cells lack a directional preference, {\it i.e.}, there is no chemotaxis, were previously experimentally observed in dendritic cells, and referred to as ``equipotential'' regions~\cite{ricart2011dendritic}. 
In these experiments, cells placed between two opposing linear gradients migrated to an intermediate position where the chemotactic response was equalized, resulting in the formation of an equipotential line.  
In the case of two point sources, the radial symmetry of the fields ensures that chemotactic responses can cancel out only along the line connecting the sources. This occurs specifically in the region between them, forming either an equipotential point or, as described above for $\lambda = \lambda^*$, an equipotential segment.
To characterize these regions, we define a direction function along the $x$-axis that indicates whether cells chemotax on average to the left or the right:
\begin{equation}\label{eq:f_x}
    f(x) = -\mathrm{sgn}(x - x_A)\,\mathrm{SNR}_A(x) - \mathrm{sgn}(x - x_B)\,\mathrm{SNR}_B(x),
\end{equation}
where $\mathrm{sgn}(x)$ denotes the sign function. Cells chemotax to the right when $f(x) > 0$ and to the left when $f(x) < 0$. The equipotential point is defined by $f(x) = 0$, which occurs only between the two sources when $\mathrm{SNR}_A(x) = \mathrm{SNR}_B(x)$.

It is useful to draw an analogy between cells within the chemical sources and an overdamped particle moving under a potential $\Phi$. 
In this analogy, the dynamics are described by $\dot{x} = -\nabla \Phi(x)$, where the potential is given by $\Phi(x) = -\int f(x)\,dx$. 
Equipotential points correspond to locations where $\nabla \Phi(x) = 0$, which represent the extrema (crests and troughs) of $\Phi(x)$, as illustrated in Fig.~\ref{fig:figure2}\textbf{c}. 
Crests correspond to unstable equipotential points, where cells in the vicinity move away. 
Troughs, on the other hand, represent stable fixed points, which can either coincide with equipotential points or the source positions. 
This framework allows us to build bifurcation diagrams of the different chemotactic behavior of cells, as shown in Fig.~\ref{fig:figure2}\textbf{d}. (For details on constructing these bifurcation diagrams, see Appendix \ref{app:bifurcation}.)
For the parameter $\lambda$, we observe a (subcritical) pitchfork bifurcation at $\lambda = \lambda^*$, marking the transition between regimes where cells are attracted to the sources and where they cluster between them.
Below $\lambda^*$, two stable fixed points and one unstable equipotential point arise, corresponding to cells chemotaxing toward sources. The unstable equipotential point delimits the regions of attraction of each source.  
For $\lambda>\lambda^*$, cells concentrate at a stable equipotential point located between the two sources. 
Notably, for a small range above $\lambda^*$, three stable points coexist. However, in this regime, $\Phi(x)$ is nearly flat between the sources, resulting in weak chemotaxis, as shown in the center panel of Fig.~\ref{fig:figure2}\textbf{c}.

\begin{figure*}[!ht]
    \centering
    \includegraphics[width=0.9\linewidth]{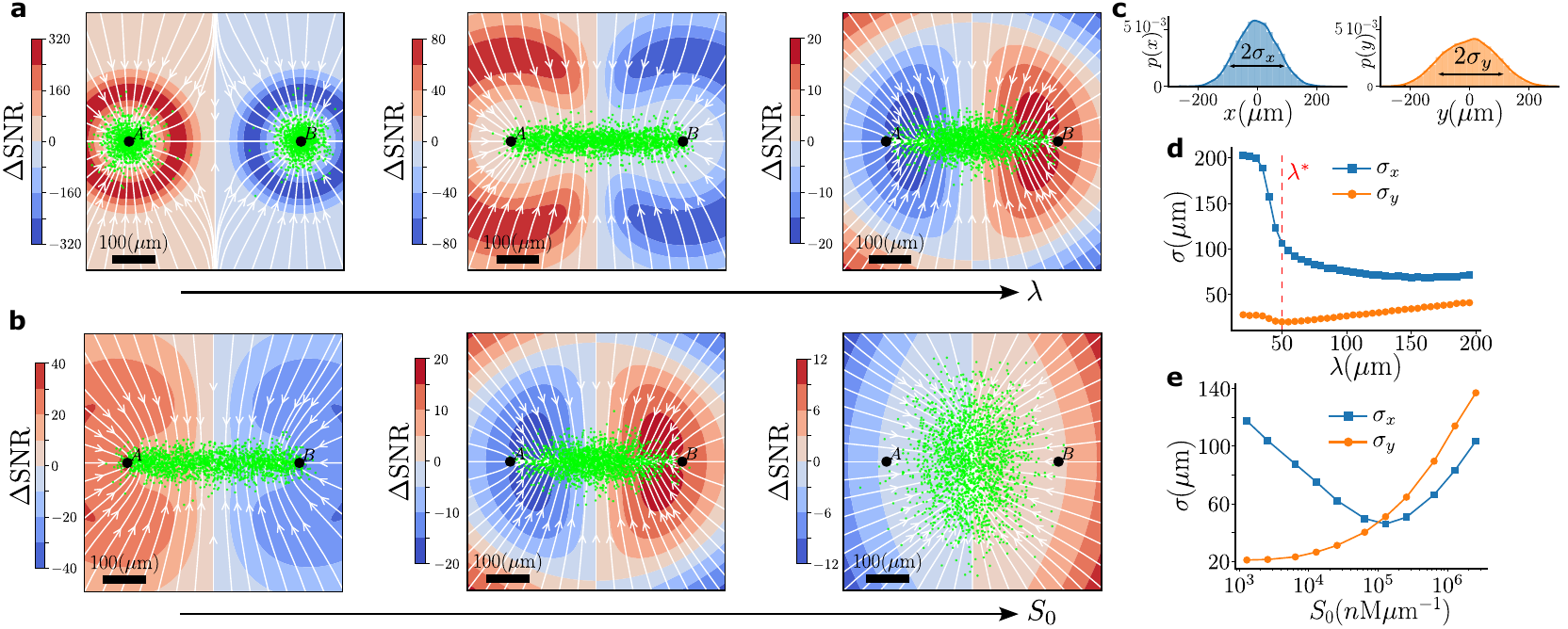}
    \caption{\textbf{(a-b)} Heatmaps of $\Delta \mathrm{SNR} = \mathrm{SNR}_A - \mathrm{SNR}_B$ across space for varying $\lambda$ (\textbf{a}) and $S_0$ (\textbf{b}). Positive values indicate greater sensing accuracy for source A's direction, while negative values indicate better sensing of source B. White streamlines represent the mean measured direction, $\bar{d}(\mathbf{r})$, and green dots show the simulated cell positions after $T=1000\,\mathrm{min}$. \textbf{(c)} Histograms of cell distributions along the $x$-axis (left) and $y$-axis (right), corresponding to the data in panel \textbf{b}-right. Cell dispersion is quantified by the standard deviation of the distributions, $\sigma_x$ and $\sigma_y$. \textbf{(d-e)} Cell dispersions, $\sigma_x$ and $\sigma_y$, as functions of $\lambda$ (\textbf{d}) and $S_0$ (\textbf{e}). Simulations parameters in (\textbf{a}): left: $\lambda=\lambda^*/2$, center: $\lambda=\lambda^*$, right: $\lambda=2\lambda^*$; in (\textbf{b}): $\lambda = 2\lambda^*$, left: $S_0=0.1S_0^*$, center: $S_0=S_0^*$, right: $S_0=50 S_0^*$. $\lambda^*=50\upmu\mathrm{m}$, $S_0^*=12810\,\mathrm{nM}\,\upmu\mathrm{m}^{-1}$. The remaining parameters are listed in Table~\ref{tab:parameters}.}
    \label{fig:figure3}
\end{figure*}

\subsection*{Complex cell distribution patterns arise from varing source properties}
To better understand how cells locally respond in the presence of two fields, we plot a heatmap representing the SNR difference, $\Delta \mathrm{SNR} = \mathrm{SNR}_A - \mathrm{SNR}_B$, across the space, as shown in Fig.~\ref{fig:figure3}. These maps highlight the regions in which the cells sense each field best and thus towards which source they are directed on average. 
For instance, cells will be directed on average towards source $A$ when $\Delta \mathrm{SNR}>0$ or to source $B$ otherwise.
Additionally, we calculate the mean estimated direction, which is just the average of Eq. \ref{eq:estimated_direction}:
\begin{equation}\label{eq:mean_estimated_dir}
    \bar{d}(\vb{r}) = \omega(\mathrm{SNR}_A)\,\hat{e}(\varphi_A) + \omega(\mathrm{SNR}_B)\,\hat{e}(\varphi_B),
\end{equation}
where $\varphi_s$ represents the true direction towards source $s$ at position $\vb{r}$, and $\omega$ is a weight function defined as $\omega = \frac{I_1}{I_0}$, with $I_n$ being the modified Bessel functions of the first kind. The function $\omega$ is monotonic and increases from 0 to 1; for high SNR values, $\omega \to 1$, while for low SNR values, $\omega \to 0$. 
For details, see Appendix~\ref{app:mean_direction}.
Thus, Eq.~(\ref{eq:mean_estimated_dir}) indicates that cells weight the direction they detect from each field according to the precision of their estimates. 
$\bar{d}(\mathbf{r})$ is a vector field that gives the average direction cells take at each spatial position; in Fig.~\ref{fig:figure3}, we plot its streamlines in white, which indicates the trajectory that cells follow on average.

The spatial extent of the distribution of cells changes with $\lambda$ (Fig.~\ref{fig:figure3}\textbf{a}). For small $\lambda$, cells form a compact, rounded cluster around the sources, exhibiting a bimodal distribution along the $x$-axis. As $\lambda$ increases, the bimodal distribution transitions to a single-peak distribution, with cells becoming confined between the two sources. 
This transition in distributions is captured by cell dispersion, $\sigma_x$, calculated as the standard deviation of the cell distribution along the $x$-axis, Fig.~\ref{fig:figure3}\textbf{c}, and plotted in Fig.~\ref{fig:figure3}\textbf{d}. For small values of $\lambda$, $2\sigma_x \approx 400 \, \upmu\mathrm{m}$, approximately the distance between the sources, reflecting cells located around sources. As $\lambda$ increases, $\sigma_x$ rapidly transitions to half its value, and the transition occurring just below the critical value $\lambda^*$. This premature transition occurs because, as  $\lambda$ approaches $\lambda^*$, $\Delta \mathrm{SNR}$ decreases to a level where random excursions allow cells to jump between the two sources. This behavior depends on the cells' persistence, characterized by the chemotactic alignment time $\tau$. Cells with longer alignment times are more persistent, requiring more time to reorient and respond to the $\Delta \mathrm{SNR}$ landscape, increasing the likelihood of switching to the other source. Conversely, for smaller $\tau$, cells can rapidly change direction in response to the $\Delta \mathrm{SNR}$ gradient, reducing the probability of switching sources. In the limit $\tau \to 0$, {we expect} the transition would become sharp at $\lambda^*$.
The cell dispersion along the $y$-axis, $\sigma_y$, initially shows a slight decrease as $\lambda$ approaches $\lambda^*$ but subsequently increases as $\lambda$ continues to grow.

We further investigate the distribution of cells for varying the concentration magnitude at the source (``source strength''), $S_0$, as shown in Fig.\ref{fig:figure3}\textbf{b}. The cell distribution transitions from being elongated along the axis connecting the two sources to becoming less elongated, with increased dispersion in the perpendicular direction. For large values of $S_0$, the anisotropy in cell dispersion reverses -- cells are more spread along the $y$-axis ($\sigma_y$ becomes larger at large $S_0$, Fig. \ref{fig:figure3}\textbf{e}).

\begin{figure*}[ht]
    \centering
    \includegraphics[width=\linewidth]{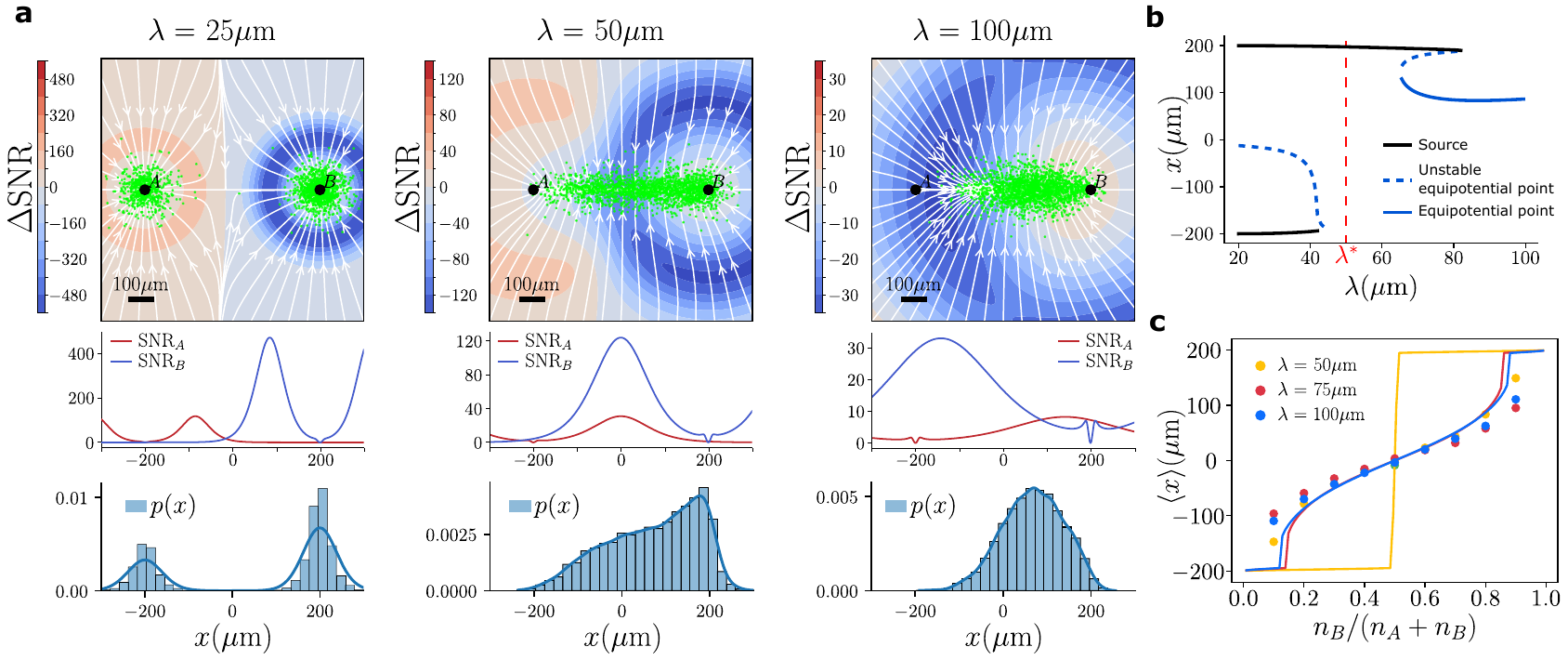}
    \caption{\textbf{(a)} (Top) Heatmap of $\Delta \mathrm{SNR}$ across space for cells with more $B$ receptors: $n_A=4000$ and $n_B=16000$. (Center) Gradient sensing SNR curves for chemoattractants $A$ and $B$, as in Fig.~\ref{fig:figure2}\textbf{b}. (Bottom) Cell dispersion along the $x$-axis. \textbf{(b)} Bifurcation diagram as a function of $\lambda$, same as Fig.~\ref{fig:figure2}\textbf{c}. The pitchfork bifurcation degenerates, creating a region where cells preferentially chemotax toward source $B$. \textbf{(c)} Stable equipotential points as a function of anisotropic sensing. Solid lines represent the theoretical equipotential points ($\mathrm{SNR}_A = \mathrm{SNR}_B$), obtained by numerically solving $\nabla \Phi(x) = 0$. Dots indicate the mean position, $\langle x \rangle$, obtained from simulations.}
    \label{fig:figure4}
\end{figure*}

\subsection*{Hierarchical chemotactic response}
Cells can exhibit differential responses to distinct chemoattractants~\cite{lundgren2023signaling}. For instance, neutrophils preferentially follow ``end-target'' chemoattractants, such as formylated peptides (e.g., fMLF), over intermediate chemoattractants secreted by host cells, such as LTB4 or IL-8. In our framework, this hierarchical response may result from variations in cellular sensitivity to each chemoattractant, e.g. differing receptor affinities ($K_{D_s}$) or receptor numbers ($n_s$).

Here, we explore cells with varying receptor numbers for each chemoattractant type (Fig.\ref{fig:figure4}). Specifically, we consider cells with $n_A = 4000$ and $n_B = 16000$ receptors, making them more sensitive to field $B$. This difference in sensitivity is captured by the signal-to-noise ratio ($\Delta \mathrm{SNR}$), resulting in an asymmetric chemotactic response map. Larger regions favor recruitment towards source $B$, as indicated by the extended blue areas and the streamlines ending at $B$ in Fig.~\ref{fig:figure4}\textbf{a}. 
When cells are initially positioned at the midpoint between the two sources, they preferentially migrate toward source $B$. After $T = 1000$ minutes, the resulting cell distributions resemble the symmetric case shown in Fig.~\ref{fig:figure3}\textbf{a}, but are noticeably skewed toward source $B$. 
Similarly, for small $\lambda$, the bimodal distribution persists, while for large $\lambda$, cells cluster around a stable equipotential point, as illustrated in the bifurcation diagram in Fig.~\ref{fig:figure4}\textbf{b}.
Additionally, we observe a shift in the equipotential points (where the SNR curves in the center panel Fig.~\ref{fig:figure4}\textbf{a} intersect): for the unstable case (small $\lambda$), the lines move toward source $A$, reflecting an expanded recruiting region of cells towards source $B$. In contrast, for the stable case (large $\lambda$), the lines shift toward source $B$, indicating cell clustering closer to this source.

An interesting situation occurs when $\lambda = \lambda^*$. Cells located to the left of source $A$ initially migrate towards source $A$ before continuing to source $B$, exhibiting multistep navigation behavior. 
Conversely, cells positioned to the right of source $B$ migrates directly to source $B$ and remains closer to it than to source $A$, emphasizing the hierarchical preference for one chemoattractant over the other.  
These behaviors mirrors that of neutrophils exposed to the ``end-target'' chemoattractant fMLP alongside ``intermediate'' chemoattractants such as IL-8 or $\mathrm{LTB}_4$~\cite{foxman1997multistep}. When initially positioned closer to the intermediate chemoattractant source, neutrophils exhibit stepwise navigation—first migrating toward the intermediate source before redirecting their trajectory toward the fMLP source. When positioned closer to end-target source, neutrophils migrated directly to this source.
The underlying mechanism for this behavior is that, in the region between the two sources, $\mathrm{SNR}_B$ is consistently larger than $\mathrm{SNR}_A$, as shown in the center panel of Fig.~\ref{fig:figure4}\textbf{a}. Consequently, while a cell may initially be attracted to source $A$, once it reaches the region between the two sources, it experiences a stronger chemotactic response to field $B$ that will eventually direct it towards source $B$. 
Notably, there is a wide range of $\lambda$ values where cells would display multistep and hierarchical navigation, as can be seen in the bifurcation diagram in Fig.~\ref{fig:figure4}\textbf{b}.

The level of asymmetry in the chemotactic response affects the position of the stable equipotential points, shown as solid lines in Fig.~\ref{fig:figure4}\textbf{c}. A larger difference in the number of receptors shifts the equipotential point closer to the source with the stronger chemotactic response, creating a more uneven cell distribution. 
Interestingly, it might seem intuitive that the mean position of the cell distribution, $\langle x \rangle$, would match the equipotential point. However, our results show that this is not always true. For $\lambda > \lambda^*$, the mean value is close to the equipotential point when the receptor difference is small. But as this difference grows, a clear gap appears, with $\langle x \rangle$ falling short of the equipotential point. 
These results highlight the need for precise metric definitions when analyzing equipotential points. The mean value, in particular, is not a reliable measure of equipotency between gradients, especially in the presence of large asymmetries. While equipotential points exhibit sharp transitions at small asymmetries, mean-based calculations tend to reflect a more gradual, continuous shift, highlighting the importance of choosing appropriate measures for capturing cell behavior.

\section*{Discussion}
In this study, we investigated how gradient sensing accuracy shapes cellular navigation in environments with multiple competing chemoattractant sources. 
In our model, the direction a cell follows during chemotaxis is determined by the accuracy of local gradient measurements.
{If cells compute a vector sum of their estimated gradient directions, this inevitably leads to a  bias in their movement toward the source with the more accurately estimated gradient.}
When both gradients are measured with similar accuracy, cells adopt an intermediate trajectory.
This simple principle recapitulates several of the non-trivial experimental observations mentioned in the introduction~\cite{foxman1997multistep,foxman1999integrating,byrne2014oscillatory,ricart2011dendritic,lin2005neutrophil,heit2002intracellular}, highlighting the fundamental role of gradient sensing accuracy in multi-gradient navigation.

In our model, SNR depends solely on fluctuations in ligand--receptor binding (Eq. \ref{eq:sensing_error}), neglecting other potential noise sources, such as fluctuations in ligand concentration, receptor number, or downstream signaling pathways. \textit{In vitro} experiments measuring SNR on various cell types in single-gradient conditions showed that receptor binding noise tends to dominate at shallow gradients or low chemoattractant concentrations, whereas intracellular noise from downstream signaling limits performance at steeper gradients and high chemoattractant concentrations~\cite{van2007biased,fuller2010external,lim2018chemotaxis,segota2013high,amselem2012control}.
Our model does include a second source of noise, independent of chemoattractant properties -- the angular noise $D_\phi$, which we will expect may be limiting when SNR is large -- but does not model downstream noise in detail. We think the core elements of our results will hold with any reasonable model of downstream noise. The key driver behind counterintuitive chemotactic behaviors—such as migration toward distant sources or confinement between them—is the non-monotonic shape of the SNR curve, and non-monotonicity is found both in experiment \cite{fuller2010external} and in models with intrinsic noise from a second messenger in the downstream signaling pathway ~\cite{ueda2007stochastic}.

Our model predicts qualitatively different behavior of cells depending on the chemoattractant environment parameters, in particular source strength ($S_0$) and span range ($\lambda$). 
We thus argue that cells can transition between different migration modes—for example, shifting from confinement between sources to directed migration—if source concentrations decrease or if sources move farther apart.  
In~\cite{byrne2014oscillatory}, neutrophils positioned between two opposing sources of LTB$_4$ and IL-8 consistently migrated toward the more distant source, despite source concentrations spanning three orders of magnitude (from nM to hundreds of nM). Similarly, in~\cite{foxman1997multistep}, neutrophils were exposed to opposing sources of LTB$_4$ and IL-8 with concentrations ranging from $\sim 1$ to $100$ nM, and they also migrated against the local gradient toward the distant source.  
These concentrations are all in excess of the relevant dissociation constants ($\sim 0.1 - 1$ nM~\cite{lin2005neutrophil, liu2020structural, yokomizo1997g}). We would then expect receptor saturation near the sources, lowering  SNR in proximity to the sources. 
Experiments with source concentrations lower than $K_D$ could help clarify the relative contribution of receptor saturation compared to other mechanisms in guiding migration toward distant sources. 
Another test of our predictions is to vary the distance between sources -- changing the separation of sources relative to the fixed $\lambda$ is akin to our modification of $\lambda$ in Fig. \ref{fig:figure2}.
Our model predicts that increasing this distance should induce a transition from confined to competing behavior between the sources.  
This transition is not apparent in earlier work~\cite{foxman1999integrating}, which tested source configurations with separations of $2\,$mm and  $1.5\,$mm. In both cases, cells migrated toward the more distant source. We expect that in this experiment, a greater source separation would be required to observe a transition to competing sources.\\

Our model relies on several key assumptions: (i) the estimates of each gradient are independent; (ii) the induced response depends solely on the signal-to-noise ratio (SNR) of the chemoattractant; and (iii) cells integrate these estimates through a vector sum. 
As noted above, chemoattractant signal integration is far more complex in living cells~\cite{ridley2003cell}.
For instance, G protein-coupled receptors (GPCRs) on the cell surface typically bind to specific chemoattractant molecules, but in some cases, a single receptor can bind multiple chemoattractants with different affinities, or conversely, a single chemoattractant can bind multiple receptors~\cite{petri2018neutrophil,metzemaekers2020neutrophil}. In human neutrophils, leukotriene B$_4$ (LTB$_4$) interacts with receptors BLT1 and BLT2, while N-formyl-methionyl-l-phenylalanine (fMLP) binds to receptors FPR1 and FPR2. Chemokines, on the other hand, exhibit extensive cross-interactions within their receptor families. In neutrophils, interleukin-8 (IL-8) binds to CXCR1 and CXCR2~\cite{petri2018neutrophil}, while in mature dendritic cells, the chemokines CCL19 and CCL21 interact with the CCR7 receptor~\cite{ricart2011dendritic}.
Once GPCRs are activated, distinct chemoattractant-receptor pairs trigger specific intracellular pathways regulated by Rho GTPases, leading to variations in signal timing and dynamic responses~\cite{ridley2003cell}. This is observed in neutrophils, where ``end-target'' chemoattractants, such as fMLP, are prioritize over ``intermediate'' signals like IL-8 and LTB$_4$, leading to hierarchical chemotactic responses~\cite{petri2018neutrophil,metzemaekers2020neutrophil,foxman1997multistep,lundgren2023signaling}.  
Additional factors—such as ligand degradation, receptor desensitization~\cite{lin2005neutrophil,lin2008modeling}, competition for receptors~\cite{dowdell2023competition} and crosstalk with other signaling pathways~\cite{byrne2014oscillatory,heit2002intracellular}—further modulate their behavior.
As a result, real physiological chemotactic responses are far more intricate than the simplified assumptions in our model, reflecting a dynamic and context-dependent integration of multiple signals. 
Additionally, experimental and theoretical studies show that integration of multiple cues is often more complex than straightforward summation in both bacteria~\cite{strauss1995analysis} and eukaryotes. 
In eukaryotic cells, integration between chemotactic and fluid flow cues may involve selection gates or hierarchical logic, enabling cells to prioritize or ignore cues based on context or processing capacity~\cite{moon2023cells}. Nevertheless, the vector sum model remains widely used and is supported by experiments where, for example, neutrophils migrate between competing stimuli consistent with vector addition~\cite{foxman1999integrating,strauss1995analysis}, or Chlamydomonas reinhardtii averages multiple orienting phototactic cues~\cite{raikwar2025phototactic}. Thus, although in vivo chemotactic responses are highly dynamic and context-dependent, our simplified approach captures core features of gradient sensing seen across diverse cell type.

Many of the factors mentioned above have been modeled in the past. For example, models that incorporate receptor desensitization have already been proposed as an essential mechanism to allow migration to distant sources~\cite{lin2008modeling,wu2011modeling,oelz2005multistep}. 
In~\cite{byrne2014oscillatory}, a model including signaling crosstalk through the inhibition of LTB$_4$ and IL-8 responses by fMLP was proposed to explain the oscillatory behavior of neutrophils navigating between these sources. 
However, none of these models account for fluctuations and different noise sources impacting the SNR, which limit a cell’s ability to estimate chemotactic gradients.
In our model, we could consider desensitization and internalization of receptors by adding new receptor states into the ligand-binding dynamics, which would make the number of available receptor to change over time. 
Incorporating receptor competition and signaling crosstalk into our framework would be more challenging, as it would break the assumption of receptor independence. 
All these model changes would need to be included in the signal-to-noise ratio (SNR) expressions that govern chemotactic gradient sensing accuracy. 
Different downstream pathway dynamics could be incorporated into our model by implementing a weighted vector sum, where pathways that induce stronger responses contribute with a higher weight. This approach could account for mechanisms such as the winner-takes-all strategy proposed to explain oscillatory behavior in neutrophils under opposing chemoattractant gradients~\cite{byrne2014oscillatory}. 
In addition, downstream signaling pathways may introduce new sources of noise into gradient sensing different from the associated in the receptor binding dynamics affecting the SNR.

When is multi-source chemotaxis biologically relevant? One well-supported idea is that multi-source chemotaxis enables long-distance migration and relay mechanisms. In neutrophils, migration from endothelial cells toward infection sites involves intermediate gradients of LTB$_4$ and IL-8, which are released at intermediary sites by other cells, as well as ``end-target'' gradients of fMLP and C5a, which are produced in the vicinity of infecting bacteria and to which cells respond preferentially~\cite{kolaczkowska2013neutrophil,petri2018neutrophil}. This suggests that intermediate gradients serve as extensions of recruitment signals, effectively forming a relay system in which cells navigate in a stepwise manner~\cite{byrne2014oscillatory,majumdar2014new,foxman1997multistep,foxman1999integrating}. However, our model does not generate such relay systems in multi-source environments. In the simple case of intercalated sources aligned in a given direction, our model predicts that cells will become trapped at or between sources, with only noise enabling movement between them—yet this does not constitute directional migration. This finding suggests that our minimal model, in which chemotactic responses are determined by relative SNRs between sources, is sufficient to capture down-gradient migration and confinement between sources but insufficient to explain relay. Additional mechanisms, such as those described previously, may be required~\cite{byrne2014oscillatory}. In particular, relay mechanisms appear to involve LTB$_4$ waves, which may play a crucial role in guiding cells over longer distances~\cite{majumdar2014new,strickland2024self}.

Less well explored than relay is the possibility of shaping the spatial distribution of cells. Our model demonstrates that the combination of two sources can produce a wide range of spatial arrangements, from confinement around the sources to elongated distributions between them. With additional sources, even more intricate spatial patterns can emerge. Speculatively, this mechanism could be exploited to strategically organize the distribution of immune cells into specific defined regions, enhancing their ability to concentrate defenses in specific areas.

Another biologically relevant multi-source scenario involves the presence of both attractant and repellent sources, which has been shown to allow microbes to effectively chase a moving target~\cite{bloxham2024repulsion}. While our model focuses on steady-state fixed attractant sources, incorporating repellent cues could further enrich the diversity of cell distributions and behaviors in multi-source environments or allow cells to better navigate dynamically-changing environments \cite{kashyap2024trade}.

\section*{Acknowledgement}
EPI and BAC are supported by NIH R35 GM142847. This work was carried out at the Advanced Research Computing at Hopkins (ARCH) core facility  (rockfish.jhu.edu), which is supported by the National Science Foundation (NSF) grant number OAC 1920103. We thank Wei Wang for a close reading of the draft. 

\appendix

\section{Chemoattractant Distribution}\label{app:sources}
We consider a synthesis-diffusion-degradation (SDD) model for chemoattractant molecules. The steady-state distribution of a chemoattractant released by localized sources is governed by the equation:
\begin{align}\label{eq:SDD}
    D\nabla^2 c(\vb{r}) - \gamma c(\vb{r}) + \psi(\vb{r}) = 0,
\end{align}
where $D$ is the diffusion coefficient, $\gamma$ represents the degradation rate, and $\psi(\vb{r})$ denotes the source term.

Considering a point source at $\vb{r} = \vb{0}$ with $\psi(\vb{r}) = \psi_p(\vb{r}) \equiv \mathcal{S}_0\delta(\vb{r})$, the solution to Eq.~(\ref{eq:SDD}) is:
\begin{align*}
    c_p(\vb{r}) = \frac{\mathcal{S}_0}{4\pi D r}e^{-\frac{r}{\lambda}},
\end{align*}
where $\mathcal{S}_0$ is the number of chemoattractant molecules released per unit time, $\lambda = \sqrt{D/\gamma}$ is the characteristic decay length, and $r=\vert\vb{r}\vert$. However, the delta function introduces a singularity at the source, which is not a realistic description of a biological source.

To address this issue, we introduce a regularized source function:
\begin{align*}
    \psi_R(r) = \frac{\mathcal{S}_0}{(2\pi\epsilon)^{3/2}} e^{-\frac{r^2}{2\epsilon}},
\end{align*}
where $\epsilon$ represents the effective size of the source.
In this case, we solve Eq.~(\ref{eq:SDD}) using the Green's function method:
\begin{align}\label{eq:green_function_sol}
    c(\vb{r}) = \frac{1}{D}\int d^3r' G(\vb{r},\vb{r}')\psi_R(\vb{r}'),
\end{align}
where $G(\vb{r},\vb{r}')$ is the Green's function, which satisfies $\left[\nabla^2 - \lambda^{-2}\right] G(\vb{r}) = -\delta(\vb{r})$. In three dimensions, the Green's function takes the form $G(\vb{r}) = \frac{1}{4\pi \lvert \vb{r} \rvert} \exp\left(-\frac{\lvert \vb{r} \rvert}{\lambda}\right)$.
Substituting it into Eq.~(\ref{eq:green_function_sol}), we obtain:
\begin{align*}
    c(\vb{r}) &= \frac{1}{4\pi D}\int d^3r' \frac{e^{-\frac{\vert \vb{r}-\vb{r}'\vert}{\lambda}}}{\vert \vb{r}-\vb{r}'\vert} \psi_R(\vb{r}').
\end{align*}

After performing the integration, we arrive at the expression:
\begin{align}\label{eq:c_vs_r_app}
    c(\vb{r}) = S_0 \frac{e^{-r/\lambda}}{r} f(r,\epsilon),
\end{align}
where we redefine $S_0 = \frac{\mathcal{S}_0}{4\pi D}$ and the regularization factor $f(r,\epsilon)$ is given by:
\begin{align*}
    f(r,\epsilon) &= \frac{1}{2} e^{\frac{\epsilon}{2\lambda^2}} \left[ 1+\mathrm{Erf}\left( \frac{r}{\sqrt{2\epsilon}} - \sqrt{\frac{\epsilon}{2\lambda^2}}\right) \right. \\
    & \quad \left. - e^{\frac{2r}{\lambda}}\mathrm{Erfc}\left(\frac{r}{\sqrt{2\epsilon}} + \sqrt{\frac{\epsilon}{2\lambda^2}}\right)\right],
\end{align*}
where $\mathrm{Erf}(z) = \frac{2}{\sqrt{\pi}} \int_0^z e^{-t^2}dt$ is the error function, and $\mathrm{Erfc}(z) = 1 - \mathrm{Erf}(z)$ is the complementary error function.

Note, that in the limit $\epsilon \to 0$, we recover the point-source case: $\mathrm{Erf}\left(\frac{r-\epsilon/\lambda}{\sqrt{2\epsilon}}\right) \to 1$, and thus,
\begin{align*}
    c(r) \to S_0 \frac{e^{-r/\lambda}}{r}.
\end{align*}

\section{Mean estimated direction computation}\label{app:mean_direction}
Cells estimate the sensed chemotactic direction $\hat{\vb{d}}(\vb{r},t)$ as given by Eq.~(\ref{eq:estimated_direction}). 
This is a stochastic measurement that depends on the accuracy with which cells estimate the orientation of each individual chemical gradient, which in turn depends on their spatial location. 
Here, we compute the mean chemotactic direction cells measure at each position. 

Taking the mean of Eq.~(\ref{eq:estimated_direction}) and recalling that the estimates of each gradient are independent, we obtain:
\begin{align*}
    \bar{\vb{d}}(\vb{r}) &= \langle \hat{\vb{d}}(\vb{r},t) \rangle =  \langle \hat{e}_A(\vb{r},t) + \hat{e}_B(\vb{r},t) \rangle\\
    &= \langle \hat{e}_A(\vb{r},t)\rangle + \langle \hat{e}_B(\vb{r},t) \rangle.
\end{align*}

Since cells sense both gradient by the same gradient sensing mechanism,  these two averages will have the same form. Thus, we only need to compute the mean value of a single estimate:
\begin{align*}
    \langle \hat{e}_s(\vb{r},t)\rangle = \left\langle \cos\left(\phi_s(\vb{r},t)\right)\right\rangle\hat{i} + \left\langle \sin\left(\phi_s(\vb{r},t)\right)\right\rangle\hat{j}.
\end{align*}
Here, $\phi_s$ represents the estimated gradient orientation at position $\vb{r}$ and time $t$, while $s$ indexes the chemoattractant gradients and $\hat{i}$ and $\hat{j}$ are the unit vectors in the $x$ and $y$ direction, respectively.
Next, we evaluate the expected values:
\begin{align*}
    \left\langle \cos\left(\phi_s(\vb{r},t)\right)\right\rangle &= \int d\phi' \cos(\phi') p(\phi'|\varphi_s,\kappa_s),\\
    \left\langle \sin\left(\phi_s(\vb{r},t)\right)\right\rangle &= \int d\phi' \sin(\phi') p(\phi'|\varphi_s,\kappa_s).
\end{align*}
where $p(\phi|\varphi_s,\kappa_s) = \frac{1}{2\pi I_0(\kappa_s)} \exp \left(\kappa_s\cos(\phi - \varphi_s)\right)$ is the von Mises distribution, centered at $\varphi_s$—the true direction toward the gradient source—with concentration parameter $\kappa_s=\mathrm{SNR}_s$.

We start by computing the mean cosine integral:
\begin{align*}
    \left\langle \cos\left(\phi_s(\vb{r},t)\right)\right\rangle &= \int_{-\pi}^{\pi} d\phi' \cos(\phi') p(\phi'|\varphi_s,\kappa_s)\\
    &= \frac{1}{2\pi I_0(\kappa_s)} \times \\
    & \quad\int_{-\pi}^{\pi} d\phi' \cos(\phi') \exp \left(\kappa_s\cos(\phi' - \varphi_s)\right).
\end{align*}

Using the variable substitution $\theta = \phi' - \varphi_s$ and leveraging the periodicity of $p(\phi|\varphi_s,\kappa_s)$, we rewrite:
\begin{align*}
    \left\langle \cos\left(\phi_s(\vb{r},t)\right)\right\rangle &= \frac{1}{2\pi I_0(\kappa_s)} \int_{-\pi+\varphi_s}^{\pi+\varphi_s} d\theta \cos(\theta+\varphi_s) e^{\kappa_s\cos(\theta)}\\
    &= \frac{1}{2\pi I_0(\kappa_s)} \left(\cos(\varphi_s) \int_{-\pi}^{\pi} d\theta \cos(\theta) e^{\kappa_s\cos(\theta)} \right.\\
    &\quad \left. - \sin(\varphi_s) \int_{-\pi}^{\pi} d\theta \sin(\theta) e^{\kappa_s\cos(\theta)} \right).
\end{align*}

Due to symmetry, we have:
\begin{align*}
    \int_{-\pi}^{\pi} d\theta \sin(\theta) e^{\kappa_s\cos(\theta)} &= 0, \\
    \int_{-\pi}^{\pi} d\theta \cos(\theta) e^{\kappa_s\cos(\theta)} &= 2\pi I_1(\kappa_s).
\end{align*}

Thus, we obtain:
\begin{align*}
    \left\langle \cos\left(\phi_s(\vb{r},t)\right)\right\rangle = \frac{I_1(\kappa_s)}{I_0(\kappa_s)}\cos(\varphi_s).
\end{align*}

Following the same procedure for the sine integral, we find:
\begin{align*}
    \left\langle \sin\left(\phi_s(\vb{r},t)\right)\right\rangle = \frac{I_1(\kappa_s)}{I_0(\kappa_s)}\sin(\varphi_s).
\end{align*}

Finally, the expression for the mean chemotactic direction is:
\begin{align*}
    \bar{d}(\vb{r}) = \hat{e}(\varphi_A)\frac{I_1(\kappa_A)}{I_0(\kappa_A)} + \hat{e}(\varphi_B)\frac{I_1(\kappa_B)}{I_0(\kappa_B)},
\end{align*}
where $\hat{e}\left(\varphi_s\right) = (\cos\varphi_s, \sin\varphi_s)^\mathrm{T}$.

Note that the contribution of each source to the mean direction is weighted by its SNR. In the limits, $\mathrm{SNR} \to 0$,  $\frac{I_1(\kappa_s)}{I_0(\kappa_s)} \to 0$, and $\mathrm{SNR} \to \infty$, $\frac{I_1(\kappa_s)}{I_0(\kappa_s)} \to 1$.
This indicates that, in the absence of a strong signal (low SNR), the estimated direction becomes random, while in the presence of a strong signal (high SNR), the estimated direction aligns perfectly with the true gradient orientation.

\section{Equipotential Points and Bifurcations}\label{app:bifurcation}
We explore the mathematical conditions for equipotential points, where the opposing chemotactic signals from two point sources precisely balance.  Given the vector nature of the chemotactic response and the radial symmetry of point sources, this equilibrium can only occur along the axis connecting both sources, identified as the  $x$-axis. Mathematically, equipotential points $x_c$ can be determined by the direction function, defined in Eq.~(\ref{eq:f_x}), where $f(x_c) = 0$. Recall that $f > 0$ indicates cells moving on average to the right, while $f < 0$ indicates movement to the left.
The function $f(x)$ can only be zero within the region between the two sources, i.e., $x_A < x < x_B$, where $x_A$ and $x_B$ are the $x$-coordinates of the source positions. In this case, the direction function simplifies to  
\begin{align*}
f(x) = \mathrm{SNR}_B(x) - \mathrm{SNR}_A(x).    
\end{align*}

Thus, equipotential points satisfy $f(x_c) = 0$, which corresponds to the condition $\mathrm{SNR}_A(x_c) = \mathrm{SNR}_B(x_c)$. In other words, an equipotential point is the position between the sources where both SNR values are equal.

We can further classify equipotential points as stable or unstable based on the derivative of $f(x)$. A point is stable if $\frac{df}{dx}(x_c) < 0$ and unstable if $\frac{df}{dx}(x_c) > 0$.
An interesting case arises when $\frac{df}{dx}(x_c=x^*) = 0$, which marks the transition from stable to unstable solutions—i.e., the bifurcation point $x^*$. 

\subsection{Equipotential Point Solution with Gradient Sensing Error Approximation}\label{app:sensing_error_approx}
To determine the solution of $f(x_c)=0$, we need to compute the SNR from the gradient sensing error expression in Eq.~(\ref{eq:sensing_error}). For that, we use the chemoattractant concentration function derived in Appendix~\ref{app:sources}, Eq.~(\ref{eq:c_vs_r_app}). However, the exact expressions are cumbersome, so we consider the asymptotic limit far from the source, $r \gg \epsilon$, recalling that $\epsilon \ll \lambda$. 
In this limit, the relevant approximations are, $f(r) \to e^{\frac{\epsilon}{2\lambda^2}} \approx 1$, $c(r) \to S_0 \frac{e^{-r/\lambda}}{r}$, and $p(r) \to 2R_\mathrm{cell} \left(\frac{1}{r} + \frac{1}{\lambda}\right)$.
Substituting these into the expression for $\sigma_\phi^2$, we obtain:
\begin{align*}
    \sigma_\phi^2(r) = \alpha\left(\lambda\,e^{-\frac{r}{2\lambda}} + K\,r\,e^{\frac{r}{2\lambda}}\right)^2 \frac{r}{(r + \lambda)^2},
\end{align*}
where, $\alpha = \frac{2\lambda^2}{nR^2_\mathrm{cell}K}$, $K = \frac{K_D \lambda}{S_0}$, and $r = \vert \vb{r} - \vb{r}_0\vert$, with $\vb{r_0}$ the source position.
Using this approximation, the SNR can be expressed as,
\begin{align}\label{eq:SNR_approx}
    \mathrm{SNR}(\vb{r})=&\frac{\left(1+\tilde{r}\right)^2}{\alpha\tilde{r}\left(e^{-\frac{\tilde{r}}{2}}+K \tilde{r} e^{\frac{\tilde{r}}{2}}\right)^2},
\end{align}
where, $\tilde{r}=\frac{r}{\lambda}$. 

With the approximated expression for SNR we can now write the direction function, 
\begin{align}\label{eq:f_approx}
    f(x) &= - \frac{\left(1+\tilde{x}_A\right)^2}{\alpha_A\tilde{x}_A\left(e^{-\frac{\tilde{x}_A}{2}}+K_A \tilde{x}_A e^{\frac{\tilde{x}_A}{2}}\right)^2}\\ \notag
    & \quad\quad + \frac{\left(1+\tilde{x}_B\right)^2}{\alpha_B\tilde{x}_B\left(e^{-\frac{\tilde{x}_B}{2}}+K_B \tilde{x}_B e^{\frac{\tilde{x}_B}{2}}\right)^2},
\end{align}
where $\tilde{x}_A= \frac{x-x_A}{\lambda}$, and $\tilde{x}_B= \frac{x_B-x}{\lambda}$. To find the equipotential points $x^*$, we solve,
\begin{align}\label{eq:equi_points_sol}
    \frac{\left(1+\tilde{x}_A\right)^2}{\alpha_A\tilde{x}_A\left(e^{-\frac{\tilde{x}_A}{2}}+K_A \tilde{x}_A e^{\frac{\tilde{x}_A}{2}}\right)^2}
    = \frac{\left(1+\tilde{x}_B\right)^2}{\alpha_B\tilde{x}_B\left(e^{-\frac{\tilde{x}_B}{2}}+K_B \tilde{x}_B e^{\frac{\tilde{x}_B}{2}}\right)^2}.
\end{align}
This is a transcendental equation and it is not possible to determine an analytical solution. However, if both sources are identical, \textit{i.e.} $\alpha_A=\alpha_B$, $K_A=K_B$, $\lambda_A=\lambda_B$, both sides of Eq.~(\ref{eq:equi_points_sol}) become identical except for the terms $\tilde{x}_A$ and $\tilde{x}_B$. In this symmetric case, the solution must satisfy $\tilde{x}_A=\tilde{x}_B$, leading to the equipotential point,
\begin{align*}
    x_c = \frac{x_B+x_A}{2}.
\end{align*}
Although this solution is trivial, it serves as a useful confirmation of the method. For more complex scenarios Eq.~(\ref{eq:equi_points_sol}) can be solved numerically.

\subsection{Analysis of Bifurcation Points and Parameter Choices}

To determine the parameter values at which bifurcations occur, we analyze Eq.~(\ref{eq:f_approx}). As established earlier, bifurcations take place when the derivative of $f(x^*)=0$, i.e., 

\begin{align}\label{eq:df_dx}
    \left.\frac{df}{dx}\right\rvert_{x=x^*} = \left.\frac{d\mathrm{SNR}_A}{dx}\right\rvert_{x=x^*} - \left. \frac{d\mathrm{SNR}_B}{dx}\right\rvert_{x=x^*} = 0.
\end{align}

To proceed, we first compute $\frac{d\mathrm{SNR}_s}{dx}$ for each source $s$:

\begin{align*}
    \frac{d\mathrm{SNR}_A}{dx} &= \frac{d\mathrm{SNR}_A}{d\tilde{x}_A} \frac{d\tilde{x}_A}{dx} = \frac{1}{\lambda} \frac{d\mathrm{SNR}_A}{d\tilde{x}_A}, \\
    \frac{d\mathrm{SNR}_B}{dx} &= \frac{d\mathrm{SNR}_B}{d\tilde{x}_B} \frac{d\tilde{x}_B}{dx} = -\frac{1}{\lambda} \frac{d\mathrm{SNR}_B}{d\tilde{x}_B}.
\end{align*}

Substituting these expressions into Eq.~(\ref{eq:df_dx}) leads to

\begin{align*}
    \left.\frac{d\mathrm{SNR}_A}{d\tilde{x}_A}\right\rvert_{\tilde{x}_A=\tilde{x}_A^*} + \left.\frac{d\mathrm{SNR}_B}{d\tilde{x}_B}\right\rvert_{\tilde{x}_B=\tilde{x}_B^*} = 0.
\end{align*}
\\
Next, we compute $\frac{d\mathrm{SNR}_s}{d\tilde{x}_s}$ using Eq.~(\ref{eq:SNR_approx}):

\begin{widetext}
\begin{equation}\label{eq:df_dhatx}
    \frac{d\mathrm{SNR}_s}{d\tilde{x}_s}=-\frac{e^{\tilde{x}_s} (1 + \tilde{x}_s) \left(1 + (-2 + 3 e^{\tilde{x}_s} K_s) \tilde{x}_s + (-1 + 2 e^{\tilde{x}_s} K_s) \tilde{x}_s^2 + e^{\tilde{x}_s} K_s \tilde{x}_s^3\right)}{\tilde{x}_s^2 (1 + e^{\tilde{x}_s} K_s \tilde{x}_s)^3}.
\end{equation}
\end{widetext}

{Assuming identical sources such that $K_A = K_B$, and recalling from the equipotential condition in such a case that $\tilde{x}_A = \tilde{x}_B$, Eq.~(\ref{eq:df_dhatx}) becomes identical for $s = A, B$. Consequently, the bifurcation condition reduces to
\begin{equation*}
    \left.\frac{d\,\mathrm{SNR}_s}{d\tilde{x}_s}\right|_{\tilde{x}_s = \tilde{x}_s^*} = 0,
\end{equation*}
where $s$ denotes either source $A$ or $B$. 
Working algebraically from Eq.~(\ref{eq:df_dhatx}), we obtain the explicit condition:
\begin{equation}
    1 - 2\tilde{x}_s^* - \left(\tilde{x}_s^*\right)^2 + e^{\tilde{x}_s^*} K \tilde{x}_s^* \left(3 + 2\tilde{x}_s^* + \left(\tilde{x}_s^*\right)^2\right) = 0.
\end{equation}
}
Finally, rewriting the equation considering $\tilde{x}^*_s = \frac{\Delta x}{2\lambda}$, where $\Delta x = x_B - x_A$, we obtain the bifurcation equation:
\begin{widetext}
\begin{align}\label{eq:bifurcations}
    2 ({\Delta x}^2 + 8\lambda^*\Delta x - 4{\lambda^*}^2 ) - \frac{K_D}{S_0^*} e^{\frac{\Delta x}{2\lambda^*}}\left({\Delta x}^3 + 8\lambda^*{\Delta x}^2 + 12{\lambda^*}^2\Delta x\right) = 0,
\end{align}
\end{widetext}
where $\lambda^*$ is the critical length scale and $S_0^*$ the critical strength parameter, and $K_D$ the dissociation constant.

We fix the dissociation constant at $K_D = 1\,\mathrm{nM}$, consistent with experimental observations~\cite{foxman1999integrating,ricart2011dendritic}.
We choose our other default parameters so that the bifurcation occurs at relevant length scales for our problem using  Eq.~(\ref{eq:bifurcations}). If we had a fixed value for the source strength $S_0$, we could solve Eq. \ref{eq:bifurcations} to find the critical value $\lambda^*$, or if we had a known value $\lambda$, we could find the critical source strength. Because we expect separations between sources to be on the order of tens of microns for relevant dynamics, we somewhat arbitrarily pick a decay length of $\lambda^* = 50\,\upmu\mathrm{m}$. Decay lengths on this order mean that the chemoattractant concentration will decay significantly relative to its value at the source on the scale of tens of microns. Given this value of $\lambda^*$, we can then solve Eq.~(\ref{eq:bifurcations}) for $S_0^*$. This is the source strength such that $\lambda^* = 50\,\upmu\mathrm{m}$ is the critical decay length. Choosing a different value for the source strength would change the critical length scale.

With this set of parameters, we  systematically explore the system’s behavior by varying one parameter at a time while holding the others constant. The only exception to this is in Fig.~\ref{fig:figure2}\textbf{b}, where we vary $S_0$ around $S_0^*$ but move the system away from the bifurcation by setting $\lambda = 2\lambda^*$. We also highlight that in Fig.~\ref{fig:figure4}, we use the numerical values for $\lambda$ and $S_0$ we found to set the system at the bifurcation point in the symmetric case, but because $n_A \neq n_B$, the system is no longer at a true bifurcation point. The complete list of parameter values used throughout this study is provided in Table~\ref{tab:parameters}.

We also note that Eq.~(\ref{eq:bifurcations})  serves as a stability criterion for equipotential points: a positive left-hand side indicates an unstable solution, whereas a negative value corresponds to a stable solution.

\subsection{Finding the equipotential points numerically and building the bifurcation diagrams}
To identify the equipotential points shown in Fig.~\ref{fig:figure2}\textbf{c}, we numerically computed $\Phi(x)$ and applied the {\it find\_peaks} function from \texttt{SciPy}. This approach proved more robust than locating the roots of the function $f(x)$ directly. Stable equipotential points correspond to peaks in $-\Phi(x)$, while unstable points correspond to peaks in $\Phi(x)$. Using this method, we constructed the bifurcation diagrams shown in Fig.~\ref{fig:figure2}\textbf{d} and Fig.~\ref{fig:figure4}\textbf{b} by plotting the identified equipotential points as a function of the parameter $\lambda$. As expected, the bifurcations occur at the critical values predicted by the analytical results discussed above.

\section{Numerical simulations}\label{app:num_sim}
We performed numerical simulations of cell motion in the presence of two chemoattractant sources. The equations of motion, given by Eqs.~(\ref{eq:prw_rdot})-(\ref{eq:prw_thetadot}), were integrated until time $T$, using the Euler-Maruyama method:  
\begin{align*}
    \vb{r}_{i+1} &= \vb{r}_{i} + v_0 \hat{e}(\phi_i)\Delta t,\\
    \phi_{i+1} &= \phi_{i} -\frac{1}{\tau} \sin(\phi_i - \hat{\phi})\Delta t + \sqrt{2D_\phi\Delta t} \, \mathcal{N}_\theta,
\end{align*}
where $i$ indexes discrete time steps, $t = i\Delta t$, with $\Delta t$ as the simulation time step. We define $\vb{r}_{i} \equiv \vb{r}(t=i\Delta t)$ and $\phi_i \equiv \phi(t=i\Delta t)$. The term $\mathcal{N}_\theta$ is a normally distributed random number with zero mean and unit variance.

Cells update their estimated chemotactic direction every $\tau_\mathrm{sensing}$. At each update, we compute $\mathrm{SNR}_s(\vb{r}_i)$ using Eqs.~(\ref{eq:c_vs_r}) and (\ref{eq:sensing_error}) and sample $\hat{\phi}_s$ from a von Mises distribution centered at the true gradient direction,
$\phi_s(\vb{r}_i)$, with concentration parameter $\kappa_s = \mathrm{SNR}_s(\vb{r}_i)$. The true gradient direction is computed as
\begin{equation*}
    \phi_s(\vb{r}_i) = \mathrm{atan2}\left(y_s - y_i,\, x_s - x_i\right),
\end{equation*}
where $(x_i, y_i)$ and $(x_s, y_s)$ are the Cartesian coordinates of the cell position and the source $s$, respectively.

All simulation and model parameters are listed in Table~\ref{tab:parameters}.

\subsection*{Orientational noise $D_\phi$ and relaxation time $\tau$. Chemotactic index}\label{sec:CI}
A key metric for quantifying how effectively cells navigate chemical gradients is the chemotactic index (CI). In our model, CI is influenced by multiple noise sources and depends not only on the signal-to-noise ratio (SNR) but also on orientational noise, characterized by the diffusion coefficient $D_\phi$, and the cellular response time $\tau$ to chemical signals. This multi-parameter approach allows us to tune the model to closely match experimental CI values.

The SNR in our model arises from both cellular and chemical source parameters. The response time $\tau$, which quantifies the persistence of cell movement directionality, is directly obtained from experimental data. To align simulated CI with empirical observations, we systematically adjust the orientational noise level $D_\phi$.

To validate our approach, we performed simulations with $N=1000$ cells placed at a fixed distance $r_0=100\,\upmu\mathrm{m}$ from the chemical source, ensuring a constant SNR and angular distribution. The simulation ran for $100\,\mathrm{min}$, after which CI was computed as

\begin{align*}
    \mathrm{CI} = \frac{1}{N} \sum_{i=1}^{N} \cos(\phi_i).
\end{align*}

Using an experimental CI value of $0.5$ for neutrophils and an alignment time $\tau=2\,\mathrm{min}$~\cite{foxman1999integrating}, we calibrated the orientational diffusion coefficient to $D_\phi=0.4 \,\mathrm{s}^{-1}$, effectively bridging our computational model with experimental data.

\begin{table}[]
    \centering
    \begin{tabular}{p{0.3\linewidth}p{0.3\linewidth}p{0.3\linewidth}}
    \toprule
    parameter & value & source \\
    \midrule
    \midrule
    \multicolumn{3}{l}{Cell} \\
    $R_\mathrm{cell}$  & $10\,\upmu\mathrm{m}$ & \cite{ipina2022collective}\\
    $v_0$  & $5\,\upmu\mathrm{m}\,\mathrm{min}^{-1}$\ & \cite{byrne2014oscillatory,parr2019simulation}\\
    $D_\theta$  & $0.4\,\mathrm{rad}\,\mathrm{min}^{-1}$ & Appendix~\ref{sec:CI}\\
    $\tau$  & $2\,\mathrm{min}$ & \cite{foxman1999integrating}\\
    $\tau_\mathrm{sensing}$ & $1\,\mathrm{min}$ & \cite{ipina2022collective}\\
    \midrule
    \multicolumn{3}{l}{Sources} \\
    $n$  & $10000$ & \cite{ipina2022collective}\\
    $K_{D}$ & $1\,\mathrm{nM}$ & \cite{foxman1999integrating,ricart2011dendritic}\\
    $\lambda^*$ & $50\,\upmu\mathrm{m}$ & Eq.~(\ref{eq:bifurcations})\\
    $S_0^*$  & $12810\,\mathrm{nM}\,\upmu\mathrm{m}^{-1}$ & Eq.~(\ref{eq:bifurcations})\\
    \midrule
    \multicolumn{3}{l}{Simulation} \\
    $N_\mathrm{traj}$  & $10000$ \\
    $T$  & $1000\,\mathrm{min}$\\
    $\Delta t$  & $0.01\,\mathrm{min}$\\
    $\epsilon$  & $16\,\upmu\mathrm{m}^2$\\
    \bottomrule
    \end{tabular}
    \caption{Model parameters.}
    \label{tab:parameters}
\end{table}

\bibliographystyle{unsrt}
\bibliography{references}

\end{document}